\documentclass[preprint,authoryear]{elsarticle}

\usepackage{amsmath,amssymb,amstext,amscd,bm}
\journal{New Astronomy}

\begin{document}
\begin{frontmatter}

\title{Kinematics of Stellar Populations with RAVE data}

\author[istanbul]{Y\"uksel  Karata\c{s}\fnref{1}},
\author[heidelberg,wuerzburg]{Rainer J. Klement\corref{cor1}\fnref{2}}
\fntext[1]{E-mail: karatas@istanbul.edu.tr}
\fntext[2]{E-mail: klement\_r@klinik.uni-wuerzburg.de}
\cortext[cor1]{Corresponding author}

\address[istanbul]{{\.I}stanbul University, Science Faculty, Department 
of Astronomy and Space Sciences, 34119 University, {\.I}stanbul, Turkey}

\address[heidelberg]{Max-Planck-Institut f\"ur Astronomie, K{\"o}nigstuhl 17, D-69117, Heidelberg, Germany}
\address[wuerzburg]{University of W{\"u}rzburg, Department of Radiation Oncology, , D-97080 W{\"u}rzburg, Germany}

\begin{abstract}
We study the kinematics of the Galactic thin and thick disk populations using
stars from the RAVE survey's second data release together with distance estimates from 
Breddels et al. (2009). The velocity distribution exhibits the expected moving groups present in the solar neighborhood.
We separate thick and thin disk stars by applying the $X$ (stellar-population) 
criterion of Schuster et al. (1993), which takes into account
both kinematic and metallicity information. For 1906 thin disk and 110 thick disk stars classified in this way,
we find a vertical velocity dispersion, mean rotational
velocity and mean orbital eccentricity of
$(\sigma_{\rm W},\langle V_{\Phi} \rangle,\langle e \rangle)_\text{thin}$ = (18$\pm$0.3 km s$^{-1}$, 223$\pm$0.4 km s$^{-1}$, 0.07$\pm$0.07) and $(\sigma_{\rm W},\langle V_{\Phi} \rangle,\langle e\rangle)_\text{thick}$ = (35$\pm$2 km s$^{-1}$, 163$\pm$2 km s$^{-1}$, 0.31$\pm$0.16), respectively. From the radial Jeans equation, we derive a thick disk scale length in the range $1.5-2.2$ kpc, whose greatest uncertainty lies in the adopted form of the underlying potential.
The shape of the orbital eccentricity distribution indicates that
the thick disk stars in our sample most likely formed \textit{in situ} with minor gas-rich mergers and/or radial migration being the most likely cause for their orbits.
We further obtain mean metal abundances of 
$\langle\rm{[M/H]}\rangle_\text{thin} = +0.03 \pm 0.17$, and $\langle\rm{[M/H]}\rangle_\text{thick} = -0.51 \pm 0.23$, in good agreement with previous estimates.  
We estimate a radial metallicity gradient in the thin disk of -0.07 dex kpc$^{-1}$, which is larger than predicted by chemical evolution models where the disk grows inside-out from infalling gas. It is, however, consistent with models where significant migration of stars shapes the chemical signature of the disk, implying that radial migration might play at least part of a role
in the thick disk's  formation.
\end{abstract}

\end{frontmatter}

\section{Introduction\label{sec:s1}}

The Radial Velocity Experiment (RAVE) is a spectroscopic survey to measure
radial velocities and stellar atmospheric parameters ($T_\text{eff}$, $\log g$, [M/H])
of up to one million stars using the Six Degree Field 
multi object spectrograph on the 1.2 m UK Schmidt Telescope of the 
Anglo--Australian Observatory \citep{Steinmetzetal08,zwi08}. 
For most of the stars, RAVE also contains 2MASS $JHK$ photometry, 
USNO-B proper motions ($\mu_{\alpha}$, $\mu_{\delta}$),
and estimates of [$\alpha$/Fe], although the latter possess great uncertainties 
\citep[typically 0.15 dex,][]{zwi08}. 

Using empirical relations between $M_{K}$ and $(J-K)$ of main-sequence and giant 
stars from \textit{Hipparcos}, \citet{Veltzetal} have studied the kinematics of 
the thin and thick disks based on G and K type stars in RAVE.
For this kind of work, distances are of great importance in order to derive 3D space velocities. 
\citet{kle08} took a similar approach and derived absolute magnitude relations
between $M_{V_{T}}$ and $(V_{T}- H)$, where $V_T$ denotes the magnitude in the 
Tycho-2 $V$ band. They assumed that the majority of RAVE stars are 
main-sequence stars with solar metallicities, which has later been shown to be an 
invalid assumption \citep{Seabrokeetal}. The RAVE survey contains 
a large fraction of giants, for which photometric parallaxes can not be estimated.
Therefore, \citet{bre10} published an important method
to derive distances from the $(J-K)$ color and the astrophysical parameters by finding
the closest match of each star to a set of theoretical isochrones. 
Due to the limitations of their theoretical isochrone grids, the distances provided 
by Breddels et al. are only reliable for stars with {\bf $\log g >3$}.

In this paper we use a sample of $\sim4000$ putative main-sequence stars ($\log g>3$) 
from the second RAVE data release and classify the Galactic disk populations according 
to the $X$ stellar population parameter defined by \citet{sch93}. 
The $X$ parameter utilizes both kinematic and metallicity information.
This paper is organized as follows: \S~\ref{sec:s2} describes the data, distances and selection 
criteria. Our procedure to calculate the orbital parameters of our sample is 
given in \S~\ref{sec:s3}; the thin/thick disk classification and their kinematic
and chemical properties are presented in \S~\ref{sec:s4}; our conclusions are summarized in \S~\ref{sec:s5}.

\section{Data, distances and selection criteria\label{sec:s2}}

The second RAVE data release (DR2) allows us to derive relations 
between kinematics and elemental abundances in the Galactic disk components.  
We use a sample of 9696 putative main sequence stars from DR2, which have been 
separated from subgiants and giants by fitting Gaussian mixture models to the 
distribution of spectroscopic $\log g$ estimates \citep[for more details, see][]{Klementetal10}.  
The $\log g$ values of these stars range from 
$3.21\leq\log g<5.0$ for stars with $(J-K)>0.5$ (roughly K and M stars) and from 
$3.09\leq\log g<5.0$ for the bluer stars. We adopt the distance estimates 
from \citet{bre10}, because they show no evidence for clear systematic 
errors in these ranges of $\log g$ values. 

We prepare our sample by putting constraints on the errors of distance, radial velocity 
and each of the proper motion components. Based on Fig.~1 of \citet{Steinmetzetal08} 
and the work of \citet{bre10}, we choose to
remove all stars with radial velocity and proper motion component errors 
greater than 5 km s$^{-1}$ and 6 mas yr$^{-1}$, respectively. We further remove all 
stars with relative distance errors larger than 40\%.
After these cuts, there remain 4027 putative main sequence stars
for studying the kinematic and chemical differences between the thin and thick disks.
Our stars span a range of $-1.53 \leq [M/H]\leq +0.49$ in metal abundances;
the [$\alpha$/Fe] values are limited to the range $[0, +0.40]$ and have large recovery errors 
of typically 0.15 dex, because the theoretical template spectra used to derive [$\alpha$/Fe] 
are limited to only two grid points at [$\alpha$/Fe]=0.0 and $+0.40$ \citep{zwi08}.
Although RAVE is not able to measure the $\alpha$ abundances of individual stars accurately, 
the assignment of [$\alpha$/Fe] is not random, and it might be interesting to 
look at average trends of [$\alpha$/Fe] for the different stellar populations.
The galactic coordinates of our 4027 sample stars are in the ranges of  
$26^\circ \leq \vert b\vert \leq 88^\circ$ and  
$3^\circ \leq l \leq 355^\circ$.

\section{Galactic Space Velocities and Orbital Parameters\label{sec:s3}}

The Galactic space velocity components $(U,V,W)$ were computed
by applying the algorithms of
\citet{JohnsonSoderblom} to the basic observables of our 4027 
RAVE main-sequence stars: celestial coordinates
($\alpha$, $\delta$), proper motion components ($\mu_{\alpha}$,
$\mu_{\delta}$), the radial velocity $(V_\text{rad})$ and the parallax 
$\pi$. The transformation matrices given in \citet{JohnsonSoderblom} have been updated to epoch J2000 \citep{mur89}. 
We adopt a right-handed coordinate system with the $(x,y,z)$-axes pointing towards 
the Galactic center ($l = 0^\circ, b = 0^\circ$),
the direction of Galactic rotation ($l=90^\circ, b=0^\circ$) and the North 
Galactic Pole ($b=90^\circ$), respectively. $(U,V,W)$ are the corresponding Cartesian
components of a star's velocity vector with respect to the Local Standard 
of Rest (LSR). Thereby, we corrected for a solar motion 
with respect to the LSR of $(U,V,W)_{\odot} = 
(+7.5, +13.5, +6.8)$ km s$^{-1}$ \citep{FrancisAnderson}.

We have computed the peri- and apogalactic distances $(R_\text{min},~R_\text{max})$
of our sample stars by integrating each star's orbit for 3 Gyr in a Galactic potential consisting of a Miyamoto-Nagai disk \citep{miy75},
a Hernquist bulge \citep{her90} and a logarithmic spherically symmetric halo potential \citep[e.g.][]{joh99}:
\begin{equation}\begin{split}\label{eq:potential}
\Phi_\text{disk}(r) &= -\frac{GM_\text{disk}}{\sqrt{x^2+y^2+(a+\sqrt{z^2+b^2})^2}}\\
\Phi_\text{bulge}(r) &= -\frac{GM_\text{bulge}}{r+c}\\
\Phi_\text{halo}(r) &=0.5\,v_0^2\,\ln(r^2+d^2).
\end{split}
\end{equation}
Here, $r=\sqrt{x^2+y^2+z^2}$, and the potential parameters have been chosen in order to provide a nearly flat rotation curve with circular velocity $v_c\approx220$ km s$^{-1}$ at the sun (see also Section~\ref{sec:s43} below):
$M_\text{disk}= 1.0\times10^{11}M_\odot$, $M_\text{bulge}= 3.4\times10^{10}M_\odot $, $a=6.5$ kpc, $b= 0.26$ kpc, $c=0.7$ kpc, $d=12.0$ kpc and $v_0=180.0$ km s$^{-1}$.
For the integration, we used a leap-frog algorithm and 100,000 time steps, which lead to an energy conservation better than $10^{-8}$ for most stars. Each star's $(U,V,W)$ velocity components have been transformed into a Galactocentric restframe 
by adding a LSR velocity of 220\,km\,s$^{-1}$ \citep{GKF79} to $V$.
The orbital eccentricity is given by the relation 
$e = (R_\text{max}-R_\text{min})/(R_\text{max}+R_\text{min})$. Here, $R_\text{max}$ and $R_\text{min}$ are apogalactic and perigalactic distances, respectively.
For each star, we also compute its mean Galactocentric distance, or mean 
orbital radius, $R_{m}$, as the mean of apogalactic and perigalactic distances.
Contrary to $R_\text{min}$ and $R_\text{max}$, $R_m$ remains fairly robust against isotropic 
diffusion of stellar orbits \citep{Grenon87}, which is also referred to 
as ``blurring'' \citep{sch09}. This becomes important when investigating 
the evidence for an intrinsic radial metallicity gradient in the thin disk.

\begin{figure}
\centering
\includegraphics*[width = 8cm, height = 11cm]{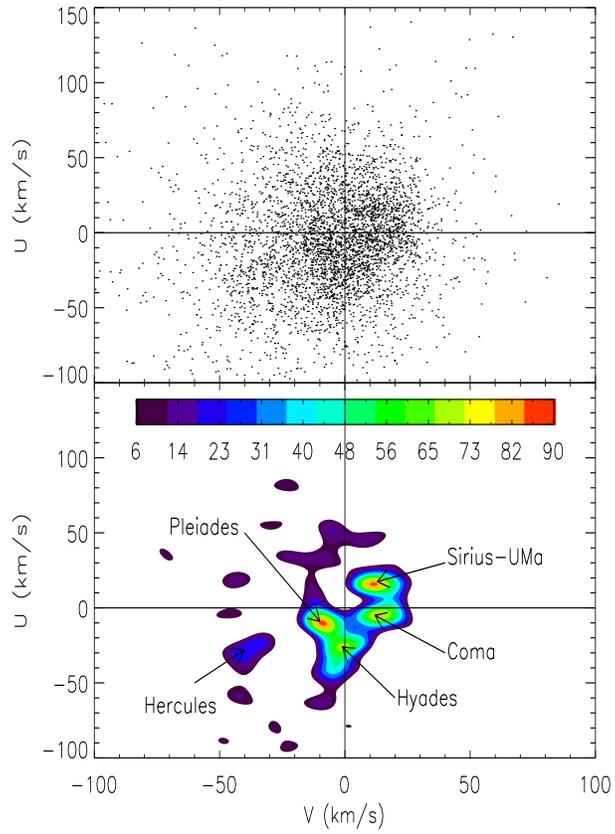}
\caption{Upper panel: The Bottlinger diagram for our sample of 4027 RAVE dwarfs. 
The displayed velocity components refer to the LSR. Bottom panel: Contours of the wavelet transform of the distribution of our data in $(U,V)$ space. The overdensities correspond to well-known moving groups in the solar neighborhood. The color bar shows the values of the wavelet coefficients that follow from eq.~\eqref{eq:wavelet}.}\label{fig:f1}
\end{figure}

The distribution of the Galactic heliocentric velocities in the $(U,V)$ 
plane is shown as a Bottlinger diagram in the upper panel of Fig.~\ref{fig:f1}. The stars appear not to be smoothly distributed as expected from the presence of moving groups.
In order to enhance any concentration of stars, we use the wavelet transform analysis, which results in a visual representation
of the stellar density in the Bottlinger diagram. We construct a quadratic grid with pixels of 1 km s$^{-1}$ width on each side and compute the wavelet coefficients at each grid point through
\begin{equation}\begin{split}\label{eq:wavelet}
w(x,y)&=\int\int dx'dy'\Psi(x-x',y-y')\times\\
	& \sum_{i=1}^{N}\delta(x'-x_i)\delta(y'-y_i)\\
	&=\sum_{i=1}^{N}\Psi(x-x_i,y-y_i),
	\end{split}\end{equation}
where $N=4027$ is the number of stars in our sample. For the analyzing wavelet $\Psi$, we use the so-called Mexican hat, which is the negative normalized second derivative of a bivariate Gaussian:
\begin{equation}
\Psi(x,y)=\bigl(2-\frac{x^2+y^2}{\sigma^2}\bigr)\,e^{-(x^2+y^2)/(2\sigma^2)}.
\end{equation} 
This type of analyzing wavelet has been used often in order to enhance overdensities in velocity diagrams \citep{sku99,ari06,kle08,zhao09}. It is normalized such that the total area under its curve is zero, so that any overdense bins will have a positive wavelet coefficient. The wavelet coefficient in such a bin will have a maximum value if the scale parameter $\sigma$ is equal to the half-width of the overdensity, assuming it's `bump' is of Gaussian shape. We expect the extend of the overdensities associated with moving groups to lie in the range 3--6 km s$^{-1}$ \citep{zhao09} and thus to be dominated by our typical velocity uncertainties that are somewhat larger. For the scale parameter $\sigma$, we therefore chose 7 km s$^{-1}$, which is of the order of the median $U$ and $V$ uncertainties. The median is more representative for the typical velocity uncertainty than the mean because it is more robust against any extreme outliers. From the bottom panel of Figure~\ref{fig:f1}, we find the main moving groups expected in a solar neighborhood sample \citep[see][and references therein]{zhao09}. Note that the location of many moving groups in velocity space is often described with respect to the sun, which we have accounted for in Figure~\ref{fig:f1}, where we plot velocities with respect to the LSR.

Like \citet{bre10}, we place a cylindrical volume on the sun with
a radius of 500 pc and a height of 600 pc. For stars in this cylinder, the velocity 
dispersions $(\sigma_{U}, \sigma_{V}, \sigma_{W})$ and mean velocities 
($\langle U\rangle$, $\langle V\rangle$, $\langle W\rangle$) are estimated 
as (37$\pm$0.4, 26$\pm$0.3, 19$\pm$0.2) and
($-$11.9$\pm$0.6, $-$20.2$\pm$0.4, $-$8.1$\pm$0.3) km\,s$^{-1}$, respectively. 
These values are almost exactly the same as the ones given in Table~1 
of \citet{bre10}, which is not surprising since our sample is 
a subset of the data used by these authors.

We further investigate how the velocity dispersions change with respect to 
metallicity. This is shown in Table~\ref{tab:t1} and Figs.~2(a)-(c), where we tabulate the velocity dispersions 
for stars in several metallicity bins. While $\sigma_U$ and $\sigma_V$ remain fairly constant 
over the sampled metallicity range, we observe a rise in $\sigma_W$ with declining metallicity. 
This rise is displayed in Fig.~2(a): $\sigma_{W}$ rises very weakly 
from $\langle\sigma_{W}\rangle \approx 15$ km s$^{-1}$ to 20 km s$^{-1}$ over the metallicity range 
$-0.50 < \text{[M/H]} < +0.50$ and begins to rise more steeply for $\text{[M/H]}<-0.50$. In the range 
$\text{[M/H]} = [-0.80, -0.50]$, $\langle\sigma_{W}\rangle$ 
is 26 km s$^{-1}$. For  $\text{[M/H]}<-0.80$, $\langle\sigma_{W}\rangle$ has increased to 40 km s$^{-1}$.
Our [M/H]--$\sigma_{W}$ relation can be compared to Fig.~10 of 
\citet{WyseGilmore95}, which shows a very similar behavior for a sample of F/G stars
drawn from the Gliese catalog and the sample of 
\citet{Edvardssonetal93}. \citet{WyseGilmore95} note that between 
[Fe/H]$=-0.40$ and $-0.50$, there is a transition from 19 to 42 km s$^{-1}$ in $\sigma_{W}$. 
Similarly, an abrupt increase in the vertical velocity 
dispersion at $\text{[Fe/H]}\sim -0.70$ has been shown by \citet{Gilmoreetal89} (their Fig.~6b) for stars taken from the samples 
of \citet{Norris87c} and \citet{Stromgren87}.

We interpret the trend observed in Figure~2(a) by considering two effects: the first is continuous heating of thin disk stars, e.g. through spiral waves, which gives the older stars higher vertical velocity dispersions. There is no guarantee, however, that older stars in our sample are indeed more metal-poor on average than the younger ones \citep[e.g.][]{hol07}. In fact, the age-metallicity relation of thin disk stars is still under debate, and it's shape for RAVE stars is currently under investigation \citep{ang09}. The weak rise of $\langle\sigma_{W}\rangle$ in the range $\text{[M/H]}=[-0.5,+0.5]$ could also be caused by a growing contribution of thick disk stars that have intrinsically higher velocity dispersions. This second effect is most probably responsible for the sharp rise of the velocity dispersion at $\text{[M/H]}<-0.80$: our value of $\sigma_W=40$ km s$^{-1}$ is too large to be explainable by disk heating alone, even if we assume a strong correlation of stellar age and metallicity \citep[see also the discussion in][Sec.~11]{hol07}.

We point out that our stars have estimates of total metallicity rather than iron abundances;
the latter are usually lower by $\sim0.1-0.3$ dex compared to [M/H] at low metallicities 
\citep[Fig.~17]{zwi08}. However,
qualitatively, our [M/H]--$\sigma_{W}$ relation is in good agreement 
with the earlier findings of \citet{WyseGilmore95} and \citet{Gilmoreetal89}.

For the metallicity ranges from Table~\ref{tab:t1}, we compute two estimates for the mean rotational 
velocity in the Galactic restframe: $\langle V_\text{rot}\rangle$ is derived on the basis of 
radial velocities and distances alone \citep{FrenkWhite}, so that (eventually systematic) errors in
the proper motions have no influence on the inferred value of rotation; thereby, 
the Galaxy is assumed to be axisymmetric. $\langle V_{\Phi}\rangle$ is based on 
the full set of observables including proper motions and is the mean rest-frame 
rotational velocity in a Galactocentric cylindrical coordinate system. In calculating 
both rotational velocities, we have assumed that the sun lies at a distance 
$R_\odot= 8.5$ kpc from the Galactic center \citep{KerrLyndenBell},
and that the velocity of the LSR is $V_\text{LSR} = 220$ km s$^{-1}$. 
As can be seen from Table~\ref{tab:t2}, both measures of rotational velocity, $ \langle V_\text{rot} 
\rangle$ and  $\langle V_{\Phi} \rangle$, remain almost constant with changing metallicity 
and are consistent with 
rotational velocities typical for the thin disk. The rotational velocities obtained 
from the algorithm of \citet{FrenkWhite} are slightly smaller than the 
$\langle V_{\Phi} \rangle$ values. This behavior has already been noticed 
by \citet{Carolloetal07}, who also found $\langle V_\text{rot} \rangle$ generally 
smaller than $\langle V_{\Phi} \rangle$ (see their Supplemental Table 2).

From the first two [M/H] bins listed in Tables~\ref{tab:t1} and \ref{tab:t2}, two likely halo stars have been excluded from 
the statistics in order not to effect the rotational velocities. Their space velocities are given separately in Table~\ref{tab:t3}.
The mean eccentricities given in Table~\ref{tab:t2} decline to more circular orbits with decreasing metallicity.
Simultaneously, $\langle [\alpha/\text{Fe}]\rangle$ values drop to solar with some fluctuations.

\begin{figure}
\centering
\includegraphics*[width = 8cm, height = 12cm]{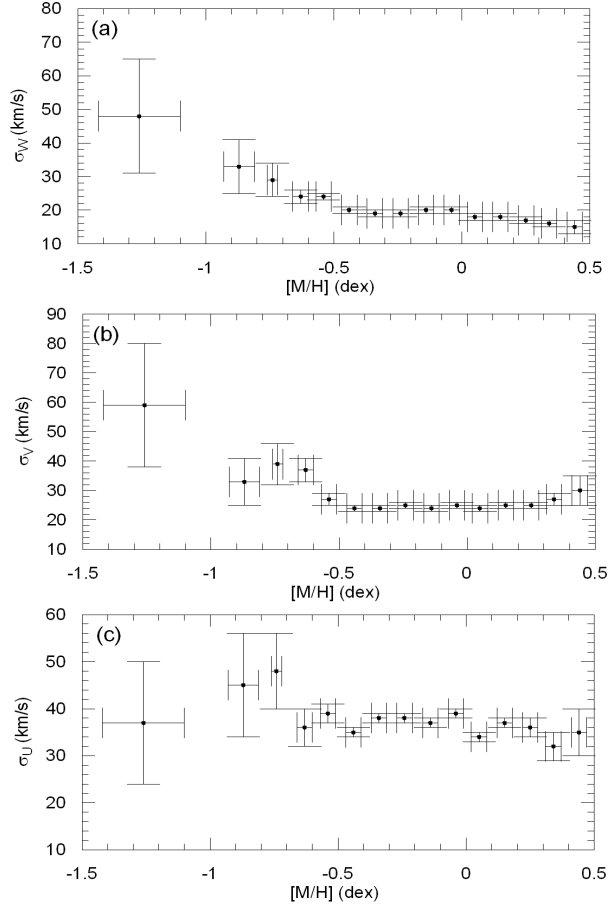}
\caption{$\sigma_{W}$, $\sigma_{V}$, $\sigma_{U}$ versus [M/H] for the data in Table~1.}\label{fig:f2}
\end{figure}

\begin{table*}
\tiny
\caption{Metallicities ([M/H], $\langle\text{[M/H]}\rangle$) and $(\sigma_{U}, \sigma_{V}, \sigma_{W})$ 
velocity dispersions for our 4027 putative main sequence stars. The covariances $\sigma^2_{UV}$, $\sigma^2_{UW}$, $\sigma^2_{VW}$
are listed in Cols.~7$-$9.}\label{tab:t1}
\begin{tabular}{ccccccccc}
\hline
[M/H] &   $ \langle\text{[M/H]}\rangle$ & $N$ &$\sigma_{U}$ &$\sigma_{V}$ &$\sigma_{W}$& $\sigma^2_{UV}$& $\sigma^2_{UW}$&$\sigma^2_{VW}$ \\ 
(dex) &(dex)&    &km s$^{-1}$&  km s$^{-1}$& km s$^{-1}$& km$^{2}$ s$^{-2}$&km$^{2}$ s$^{-2}$& km$^{2}$ s$^{-2}$  \\
\hline
[$-$1.53,$-$1.00] &$-$1.26$\pm$0.16 &  5 &37$\pm$13 &59$\pm$21 &48$\pm$17 &  1057$\pm$83&$-$115$\pm$59&$-$1356$\pm$55\\
($-$1.00,$-$0.80] &$-$0.87$\pm$0.06 &  9 &45$\pm$11 &33$\pm$8  &33$\pm$8  &   296$\pm$61&$-$153$\pm$58&    245$\pm$52\\
($-$0.80,$-$0.70] &$-$0.74$\pm$0.02 & 19 &48$\pm$8  &39$\pm$7  &29$\pm$5  &$-$424$\pm$55&$-$160$\pm$53& $-$145$\pm$45\\
($-$0.70,$-$0.60] &$-$0.63$\pm$0.03 & 48 &36$\pm$4  &37$\pm$4  &24$\pm$2  &    56$\pm$53&$-$  8$\pm$43&     57$\pm$45\\
($-$0.60,$-$0.50] &$-$0.54$\pm$0.03 &144 &39$\pm$2  &27$\pm$2  &24$\pm$1  &   166$\pm$51&   183$\pm$50&    150$\pm$40\\
($-$0.50,$-$0.40] &$-$0.44$\pm$0.03 &304 &35$\pm$1  &24$\pm$1  &20$\pm$1  &    91$\pm$45&    15$\pm$41&     70$\pm$33\\
($-$0.40,$-$0.30] &$-$0.34$\pm$0.03 &494 &38$\pm$1  &24$\pm$1  &19$\pm$1  &   161$\pm$48& $-$29$\pm$42&  $-$24$\pm$42\\
($-$0.30,$-$0.20] &$-$0.24$\pm$0.03 &581 &38$\pm$1  &25$\pm$1  &19$\pm$1  & $-$49$\pm$44& $-$35$\pm$42&     25$\pm$32\\
($-$0.20,$-$0.10] &$-$0.14$\pm$0.03 &605 &37$\pm$1  &24$\pm$1  &20$\pm$1  &   247$\pm$49& $-$10$\pm$42&  $-$17$\pm$31\\
($-$0.10,$-$0.00] &$-$0.04$\pm$0.03 &772 &39$\pm$1  &25$\pm$1  &20$\pm$1  &   106$\pm$49&    12$\pm$44&      4$\pm$26\\
($-$0.00, +0.10]  &  +0.05$\pm$0.03 &410 &34$\pm$1  &24$\pm$1  &18$\pm$1  & $-$25$\pm$41& $-$48$\pm$37&      1$\pm$30\\
(+0.10,   +0.20]  &  +0.15$\pm$0.03 &309 &37$\pm$1  &25$\pm$1  &18$\pm$1  &    41$\pm$46& $-$28$\pm$40&     20$\pm$31\\
(+0.20,   +0.30]  &  +0.25$\pm$0.03 &222 &36$\pm$2  &25$\pm$1  &17$\pm$1  &   127$\pm$47&    23$\pm$40&   $-$9$\pm$30\\
(+0.30,   +0.40]  &  +0.34$\pm$0.03 & 81 &32$\pm$3  &27$\pm$2  &16$\pm$1  &   168$\pm$46&    18$\pm$36&      1$\pm$31\\
(+0.40,   +0.50]  &  +0.44$\pm$0.03 & 22 &35$\pm$5  &30$\pm$5  &15$\pm$2  &$-$104$\pm$44& $-$90$\pm$36&    131$\pm$37\\
\hline
\end{tabular}  
\end{table*}

\begin{table*}
\tiny
\caption{Mean metallicities, rotational velocities, 
eccentricities and alpha element abundances 
for our 4027 putative main sequence stars. The uncertainties in $\langle\text{[M/H]}\rangle$, $\langle e \rangle$ and $\langle[\alpha/\text{Fe}]\rangle$ are
the standard deviations. The uncertainties in $V_\text{rot}$ and $V_{\Phi}$ are the mean errors.}\label{tab:t2}
\begin{tabular}{ccccccc}
\hline
[M/H] & $\langle\text{[M/H]}\rangle$ & $N$ & $\langle V_\text{rot}\rangle$  & $\langle V_{\Phi}\rangle$ & $\langle e \rangle$ & $\langle[\alpha/\text{Fe}]\rangle$ \\
(dex) &(dex)&    &km s$^{-1}$&km s$^{-1}$& &(dex)  \\

\hline
[$-$1.53, $-$1.00]  & $-$1.26$\pm$0.16 &    5 &    131$\pm$32&    219$\pm$26 &0.21$\pm$0.15 & +0.18$\pm$0.07 \\
($-$1.00, $-$0.80]  & $-$0.87$\pm$0.06 &    9 &    117$\pm$30&    203$\pm$11 &0.21$\pm$0.15 & +0.19$\pm$0.13 \\
($-$0.80, $-$0.70]  & $-$0.74$\pm$0.02 &   19 &    172$\pm$15&    206$\pm$9  &0.19$\pm$0.13 & +0.24$\pm$0.11 \\
($-$0.70, $-$0.60]  & $-$0.63$\pm$0.03 &   48 &    182$\pm$9 &    219$\pm$5  &0.14$\pm$0.14 & +0.25$\pm$0.10 \\
($-$0.60, $-$0.50]  & $-$0.54$\pm$0.03 &  144 &    175$\pm$5 &    215$\pm$2  &0.14$\pm$0.10 & +0.24$\pm$0.10 \\
($-$0.50, $-$0.40]  & $-$0.44$\pm$0.03 &  304 &    185$\pm$3 &    219$\pm$1  &0.13$\pm$0.08 & +0.21$\pm$0.08 \\
($-$0.40, $-$0.30]  & $-$0.34$\pm$0.03 &  494 &    183$\pm$2 &    217$\pm$1  &0.13$\pm$0.09 & +0.17$\pm$0.09 \\
($-$0.30, $-$0.20]  & $-$0.24$\pm$0.03 &  581 &    181$\pm$2 &    215$\pm$1  &0.13$\pm$0.09 & +0.13$\pm$0.08 \\
($-$0.20, $-$0.10]  & $-$0.14$\pm$0.03 &  605 &    184$\pm$2 &    214$\pm$1  &0.13$\pm$0.09 & +0.10$\pm$0.09 \\
($-$0.10, $-$0.00]  & $-$0.04$\pm$0.03 &  772 &    180$\pm$2 &    212$\pm$1  &0.14$\pm$0.09 & +0.07$\pm$0.07 \\
($-$0.00,   +0.10]  &   +0.05$\pm$0.03 &  410 &    177$\pm$2 &    212$\pm$1  &0.13$\pm$0.09 & +0.07$\pm$0.07 \\
(+0.10,     +0.20]  &   +0.15$\pm$0.03 &  309 &    177$\pm$3 &    209$\pm$1  &0.14$\pm$0.09 & +0.05$\pm$0.06 \\
(+0.20,     +0.30]  &   +0.25$\pm$0.03 &  222 &    181$\pm$3 &    211$\pm$2  &0.14$\pm$0.09 & +0.03$\pm$0.05 \\
(+0.30,     +0.40]  &   +0.34$\pm$0.03 &   81 &    175$\pm$5 &    208$\pm$3  &0.14$\pm$0.09 & +0.03$\pm$0.04 \\
(+0.40,     +0.50]  &   +0.44$\pm$0.03 &   22 &    197$\pm$12&    205$\pm$6  &0.16$\pm$0.10 & +0.01$\pm$0.05 \\
\hline
\end{tabular}  
\end{table*}

\begin{table*}
\tiny
\caption{Distances, metallicity estimates and velocity components for the two stars 
in the first two metal abundance ranges of Tables~\ref{tab:t1} and \ref{tab:t2}.}
\label{tab:t3}
\begin{tabular}{ccccccc}
\hline
OBJECT-ID       &d  & [M/H] & $U$  & $V$  &  $W$   &  $\langle V_{\Phi} \rangle$ \\
 & kpc &(dex)  &km s$^{-1}$&km s$^{-1}$&km s$^{-1}$ &km s$^{-1}$   \\
\hline
T8093-00436-1     &0.11    & $-$1.45  &   55.39 &$-$309.82 &$-$90.04 &$-$90 \\
C0229387-130746   &0.42    & $-$0.99  &$-$99.62 &$-$146.29 &$-$135.33 &  74 \\
\hline
\end{tabular}  
\end{table*}

\section{Separation of Galactic thin and thick disk populations and derivation of their chemo-kinematical properties\label{sec:s4}}

\subsection{Separating thin and thick disk stars in the ([M/H],$V$) plane\label{sec:s41}}

We wish to separate our sample into members of the thin and thick disk. On average, thin disk stars tend to be younger, more metal-rich and faster rotating than their thick-disk counterparts. Due to a lack of metallicity estimates and/\,or stellar ages, many studies have based their thin/\,thick disk separation on the stellar kinematics only \citep[e.g.][]{ben03}. However, such an approach often requires fixing some values for the velocity dispersions of the thin and thick disk components a priori and might perform suboptimal due to some overlap in the velocity components of thin and thick disk stars. We therefore aim at using both kinematics and metallicities for our thin/\,thick disk separation. 
 
We expect our sample to be dominated heavily by the thin disk, because most of the stars possess solar-like metallicities \citep[see also Sec.~4 in][]{bre10}. Due to this imbalance, we expect that any unsupervised\footnote{``Unsupervised'' meaning without having a labeled set of training data.} classification method might give non optimal results when we restrict ourselves to the RAVE data only. We have applied an expectation-maximization (EM) algorithm to the distribution of our data in the ([M/H],$V_\Phi$) plane. The EM algorithm fits a specified number of bivariate Gaussian components to the data by iteratively maximizing the likelihood of the data given the Gaussian model parameters. The EM algorithm is stopped after 100 iterations, which is sufficient to obtain robust parameter estimations. To prevent convergence to local maxima, we have first initialized the centroids of the Gaussians using a k-nearest neighbor algorithm with 100 iterations before each run. We checked that different initial guesses, that were partly motivated by the actual distribution of our data in the ([M/H],$V_\Phi$) plane, converged to the same centroids. We found that three components give a higher likelihood than one, two or four components. By connecting one of the Gaussians (say, number 1) to the thick disk component, we then classify star $i$ as belonging to the thick disk if the probability given by
\begin{equation}\label{eq:prob}
	P(\text{thick}\,\vert\,{\bm x}_i)=\frac{\pi_1\mathcal{N}\bigl({\bm x}_i\,\vert\,{\bm\mu}_1,{\bm\Sigma}_1\bigr)}{\sum_{k=1}^3\pi_k\mathcal{N}\bigl({\bm x}_i\,\vert\,{\bm\mu}_k,{\bm\Sigma}_k\bigr)}
\end{equation}
is greater than 0.5. Here,
\begin{equation}
	\bm x_i=\bigl(\begin{smallmatrix}\text{[M/H]}\\V_\Phi\end{smallmatrix}\bigr)
\end{equation}
is the ``feature vector'' of star $i$ containing its metallicity and rotational velocity component, $\pi_k$ denotes the weight of the $k$th Gaussian (with $\sum_{k=1}^3\pi_k=1$), and $\mathcal{N}\bigl({\bm x}_i\,\vert\,{\bm\mu}_k,{\bm\Sigma}_k\bigr)$ is given as
\begin{equation}
\mathcal{N}\bigl({\bm x}_i\,\vert\,{\bm\mu}_k,{\bm\Sigma}_k\bigr)=\frac{1}{2\pi\,\vert\bm\Sigma_k\vert^{1/2}}\exp\Bigl(-\frac{1}{2}(\bm x_i-\bm\mu_k)^\text{T}\bm\Sigma_k^{-1}(\bm x_i-\bm\mu_k)\Bigr).
\end{equation}
$\bm\mu_k$ and $\bm\Sigma_k$ denote the mean and ($2\times2$) covariance matrix of the $k$th Gaussian component, respectively.

\begin{figure}
\centering
\includegraphics*[width = 0.9\textwidth]{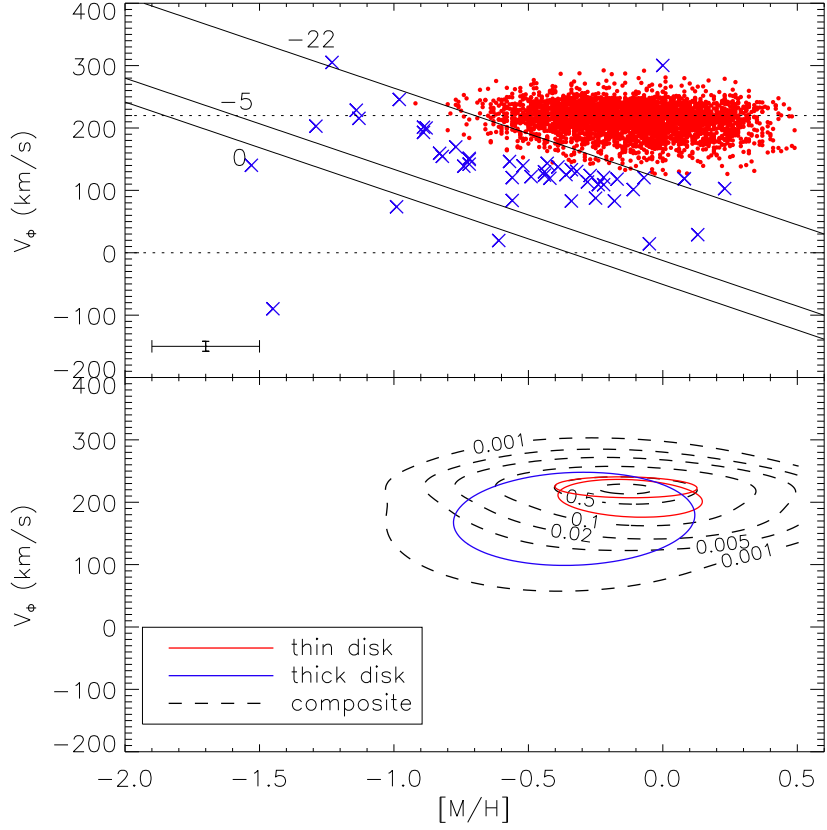}
\caption{Upper panel: The distribution of our 4027 sample stars in the ([M/H],$V_\Phi$) plane. The red dots and blue crosses denote stars classified as belonging to the thin and thick disk, respectively, according to a Gaussian mixture model with three components (see text). The solid diagonal lines are lines of constant X; the values of X are denoted for each line. Bottom panel: Dashed lines indicate contours of the three component joint probability density function according to 0.1\%, 0.5\%, 2\%, 10\%, 50\% and 90\% of the maximum. Solid lines show contours at 50\% of the maximum of each individual component. The red and blue contours correspond to the Gaussian component(s) that we associate with the thin and thick disk, respectively. The horizontal dashed lines indicate rotation velocities of $0$ and 220 km s$^{-1}$.}\label{fig:f3}
\end{figure}

The upper panel of Figure~3 shows the distribution of our data in the ([M/H],$V_\Phi$) plane; the lower panel shows contours of the three Gaussian components that we fit to these data. The solid red and blue contours correspond to 50\% of the individual maximum probability density of each Gaussian, while the dashed ones indicate various fractions of the composite probability density function. The blue Gaussian is the one we attribute to the thick disk, and the stars classified as thick disk stars according to equation~\eqref{eq:prob} are indicated as blue crosses in the upper panel of Figure~3. For comparison, we also show three straight diagonal lines that have been introduced by \citet{sch93} to separate the thin disk, thick disk and halo stars. Each line is a linear combination of [M/H] and $V_\Phi$ given by
\begin{equation}
 V_\Phi=-\frac{19.08\,\text{[M/H]}+6.63+X}{0.1305},
\end{equation}
where $X$ is the stellar population parameter defined by \citet{sch93} as the separation from the line $X=0$, which itself is defined to run through the two points ([M/H],$V_\Phi$) = (-0.3,0 km s$^{-1}$) and (-1.5,175 km s$^{-1}$). According to \citet{sch93}, the line labeled ``0'' divides the halo from the thick disk population, and stars become more disk-like with decreasing $X$. For our RAVE sample, we will classify stars as thick disk members if their location in the ([M/H],$V_\Phi$) plane satisfies $-22<X<-5.$\footnote{\citet{sch93} and \citet{sch06} adopted slightly different boundaries, $-21<X<-6$ in the ([Fe/H],$V_\Phi$) plane.} We see that most of the stars classified by our mixture model as thick disk stars are contained within this thick disk boundary, but many stars classified as thin disk members too. On the other hand, some stars are clearly misclassified by our mixture model, in particular three low metallicity stars with $V_\Phi<100$ km s$^{-1}$ and the one star at ([M/H],$V_\Phi$)=$\sim$(0,300 km s$^{-1}$). Assuming the classification based on the $X$ criterion as ``correct'' for the moment, this means a low completeness and high contamination of the thick disk sample resulting from the mixture model.

As stated before, the mixture model classification might be attenuated by the imbalance between thin and thick disk stars in our sample. We confirm this suspicion by showing that a mixture model classification of a more balanced data set yields more reliable results. For this, we choose the 1223 stars from \citet{sch06}, because they contain a considerable fraction of thick disk and halo stars. Furthermore, \citet{sch06} have shown that a histogram of the $X$ parameter for these stars can be fitted by three Gaussian distributions interpreted as thin disk, thick disk and halo (their Fig.~6). For a better comparison to our RAVE data, we have converted the [Fe/H] values of the \citeauthor{sch06} data into [M/H] by using equation~(21) from \citet{zwi08}; we also have converted their velocity components to the value of the solar motion used in this study \citep[taken from][]{FrancisAnderson}. We then classify them with the same EM algorithm we used for the RAVE sample, but now with four instead of three components\footnote{The likelihood for a four component fit is higher than for only three components.}. We attribute two of these to the halo, and the other two to the thin and thick disk, respectively. A star gets classified as belonging to a particular class if its probability for that class exceeds the probability for the other two classes. Figure~4 shows the outcome of this classification which can be compared to a separation based on the $X$ parameter. Red, blue and green crosses in the upper panel correspond to stars classified as thin disk, thick disk and halo members, respectively. In the lower panel, we compare the data from \citet{sch06} (gray crosses) to our RAVE data (black dots) in the ([M/H],$V_\Phi$) plane. For the former, typical [M/H] uncertainties at [M/H]=-0.5 are $\pm1.4$ dex \citep{sch06}, while for the latter we adopt $\pm0.2$ dex \citep{zwi08}. The larger error bars of the RAVE data could be responsible for the larger spread in the [M/H] direction observed in figure~4. In the $V_\Phi$ direction, the uncertainties of both data sets are comparably small ($\pm6.0$ km s$^{-1}$ for our RAVE data, $\pm$6.6 km s$^{-1}$ for the data of \citeauthor{sch06}). Concerning the classification of stars into the Galactic components, we see that the boundary between thin and thick disk resulting from the mixture model can be approximated by the diagonal line corresponding to $X=-22$. Also, the line $X=-5$ agrees well with the mixture model boundary between thick disk and halo in the metallicity range $-0.8\lesssim\text[M/H]\lesssim 0.4$, while there is some scattering of a few mixture model ``halo'' stars across this boundary at lower and higher metallicities (and vice versa, we find some ``thick disk'' stars in a region ascribed to the halo by the $X$ parameter). Accounting for the expected differences between the linear decision boundaries from the $X$ criterion and the more complex and curved ones from the mixture model, we conclude that both classification methods agree well. We use this fact together with the similar distribution of our RAVE stars and the thin disk stars from \citeauthor{sch06} in the ([M/H],$V_\Phi$) plane to justify the use of the $X$ parameter for classifying our sample. 

\begin{figure}
\centering
\includegraphics*[width = 0.9\textwidth]{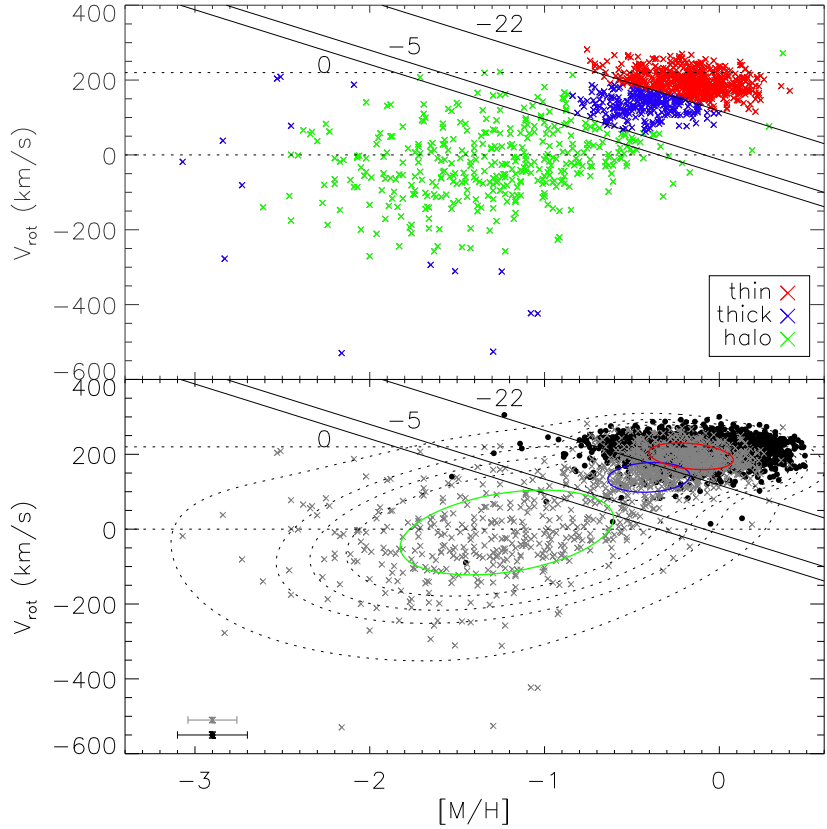}
\caption{Upper panel: The distribution of 1223 stars taken from \citet{sch93} in the ([M/H],$V_\Phi$) plane. Red, blue and green crosses denote stars classified as belonging to the thin disk, thick disk and halo, respectively. This classification is based on a Gaussian mixture model with four components. The solid diagonal lines are lines of constant X with the value of X denoted for each line. Bottom panel: Dashed lines indicate contours of the four-component joint probability density function according to 0.1\%, 0.5\%, 1\%, 2\% and 10\% of the maximum. Solid lines show contours at 50\% of the maximum of each individual class (two Gaussians for the halo). The grey crosses are the 1223 stars from \citet{sch93}, while the black dots give the location of our RAVE stars. Typical error bars of both data at [M/H]=-0.5 are given in the bottom left. Note the good agreement between our mixture model and the $X$ criterion in separating the thin and thick disk.}\label{fig:f4}
\end{figure}

\begin{table*}
\tiny
\caption{Mean metal abundances, velocity dispersions, rotational velocities 
$\langle V_\text{rot}\rangle$, $\langle V_{\Phi}\rangle$, mean eccentricity  $\langle e\rangle$, 
and alpha element abundance $\langle[\alpha/\text{Fe}]\rangle$ of thin and thick disk stars that 
are classified according to their $X$ parameter.}\label{tab:t4}
\begin{tabular}{cccccccccc}
\hline
 $X$ & $N$ & $\langle\text{[M/H]}\rangle$ & $\sigma_{U}$ & $\sigma_{V}$ & $\sigma_{W}$ & $\langle V_\text{rot}\rangle$ & $\langle V_{\Phi}\rangle$ & 
  $\langle e \rangle$ & $\langle[\alpha/\text{Fe}]\rangle$\\
\hline
 $-22<X<-5$    & 110 &$-$0.51$\pm$0.23 & 57$\pm$4 & 36$\pm$2   & 35$\pm$2   &113$\pm$7   & 163$\pm$3     & 0.31$\pm$0.16 &0.19$\pm$0.11 \\
 $X\leq-33$      &1906 &+0.03$\pm$0.17   & 34$\pm$1 & 19$\pm$0.3 & 18$\pm$0.3 &192$\pm$1   & 223$\pm$0.4 &  0.07$\pm$0.07 &0.12$\pm$0.07 \\
\hline
\end{tabular} 
\end{table*}

\subsection{Deriving the chemo-kinematical properties of thin and thick disk\label{sec:s42}}
\subsubsection{Rotational velocities}

For the 110 stars classified as members of the thick disk in the previous section, we find a mean metallicity of
$\langle\text{[M/H]}\rangle = -0.51\pm0.23$, which is in good agreement with 
the values $-0.50$ of \citet{sch93} and $-0.55$ of \citet{kar05}.
The velocity dispersions are $(\sigma_{U}, \sigma_{V}, \sigma_{W})$ = (57$\pm$4, 36$\pm$2, 35$\pm$2) 
km\,s$^{-1}$. The vertical velocity dispersion $\sigma_{W}= (35\pm$2)\,km\,s$^{-1}$ of the thick disk
agrees well with recent estimates that we summarize in Table~\ref{tab:t5}.
It also agrees well with older literature values of 30--39 km s$^{-1}$ given 
by \citet{Norris87c}, \citet{Carneyetal89} and \citet{Croswelletal}. 
 
\begin{table*}
\small
\caption{Literature values for the velocity ellipsoid of the thick disk. All values are given in units of km s$^{-1}$.}\label{tab:t5}
\begin{tabular}{ccccc}
\hline
 $\sigma_{U}$ & $\sigma_{V}$ & $\sigma_{W}$ & $\langle V_{\Phi}\rangle$ & reference\\
\hline
 50$\pm$3 & 56$\pm$3   & 34$\pm$2   & 190$\pm$5  & \citet{chi00} \\
 63$\pm$6 & 39$\pm$4   & 39$\pm$4   & 169$\pm$5  & \citet{sou03} \\
 ... & ...   & 32$\pm$5   & 154$\pm$6  & \citet{kar05} \\
 74$\pm$11 & 50$\pm$7   & 38$\pm$7   & 178$\pm$8  & \citet{val06} \\
 53$\pm$2 & 51$\pm$1   & 35$\pm$1   & 182$\pm$2  & \citet{car10} \\
 57$\pm$4 & 36$\pm$2 & 35$\pm$2 & 163$\pm$3 & this study \\
\hline
\end{tabular} 
\end{table*}

For the thin disk, we find the mean metallicity to be consistent with solar: [M/H]$=+0.03\pm0.17$.
The velocity dispersions are    
$(\sigma_{U}, \sigma_{V}, \sigma_{W})$ = (34$\pm$1, 19$\pm$0.3, 18$\pm$0.3) km s$^{-1}$.
Our estimate for the vertical velocity dispersion, $\sigma_{W} = (18\pm$0.3) km s$^{-1}$, 
is in very good agreement with the value of
(18$\pm$1) km s$^{-1}$ from \citet{Nordstrometal} and the value of (20$\pm$1) km s$^{-1}$
from \citet{sou03}.

The mean rotational velocity $\langle V_{\Phi}\rangle$, that we obtain for the thick disk, is
(163$\pm$3) km s$^{-1}$ and agrees well with literature values of
(160$\pm$30) km s$^{-1}$ \citep{Norris86} and (157$\pm$4) km s$^{-1}$ \citep{all06}, although \citet{kar05} found a slightly lower value of (154$\pm$6) km s$^{-1}$.
However, as is evident from Table~\ref{tab:t4}, the $\langle V_\text{rot}\rangle$ value of 
113$\pm$7 km s$^{-1}$, which has been calculated without using proper motions, 
is smaller than that of $\langle V_{\Phi}\rangle$ and does not agree with other estimates
from the literature. This indicates that utilizing the full available information 
is superior to replacing measured informations (proper motions) through
assumptions (an axisymmetric potential).

\subsubsection{The eccentricity distribution}
Thick disk stars in our sample display a mean eccentricity 
of $\langle e \rangle = 0.31\pm0.16$ (Fig.~5), while
for thin disk stars we find $\langle e\rangle$ = 0.07$\pm$0.07. The latter result is
fully expected, since it implies almost circular orbits for thin disk stars. 
The mean eccentricity we find for the thick disk
supports the view that accretion as a major mechanism in thick disk formation 
can be ruled out, at least for our stars at moderate heights above and below the plane.
The reason is that accreted stars would broaden and shift the eccentricity 
distribution towards higher values \citep{Salesetal}. To gain further insights, we have overplotted in Fig.~5 the predictions 
of two other thick disk formation models taken from \citet{Salesetal}, namely radial migration and minor mergers. These are eccentricity distributions
predicted for disk stars in the height range 1 kpc $\lesssim z\lesssim$ 3 kpc in order to minimize any contamination from the thin disk in these models \citep[assuming the thick disk scale height to be $\sim1$ kpc, see][Fig.~3]{Salesetal}.
Our stars are placed much closer to the plane ($z<500$ pc), but through the $X$-criterion we nevertheless are able to pick out stars from the thick disk. Although the potential used by \citet{Salesetal} to compute eccentricities is different from ours, both potentials result in a similar rotation curve near the sun and they should yield similar values of $e$ \citep[see also][]{die10}. We therefore can at least make a
qualitative comparison of our eccentricity distribution with the ones shown in \citet{Salesetal}. Such a comparison suggests that
minor mergers and/~or radial migration might be the main mechanisms responsible for placing our thick disk stars at their orbits, although the radial migration scenario used by \citet{Salesetal}
predicts a narrow peak at low eccentricity which is not observed.

We have further divided the stars formed in minor mergers into \textit{in situ} and accreted stars
and overplot the distribution of the \textit{in situ} stars in Figure~5 (again taken from \citet{Salesetal}). The shape of this distribution fits even better to our observed one as it substantially
decreases the number of stars in the high eccentricity tail (which result from accretion). Therefore, we can conclude that at least the fraction of thick disk
stars close to the plane have formed \textit{in situ}. Their $e$-distribution suggests that a majority of them might have formed \texttt{in situ} from gas that has been accreted during a period of minor mergers at high redshift \citep{bro04}. Two recent studies specifically aimed at inferring the origin of the thick disk from its eccentricity distribution have come to the same conclusions \citep{die10,wil10}.
We note, however, that care must be taken with our conclusions because they are based on a comparison of the $e$ distribution in different heights above/below the plane. Furthermore, as also pointed out by \citet{die10}, we only compare to one particular simulation for each of the thick disk formation scenarios, namely the one shown in Fig.~3 of \citet{Salesetal}. 
  
\begin{figure}
\centering
\includegraphics*[width = 8cm, height = 8cm]{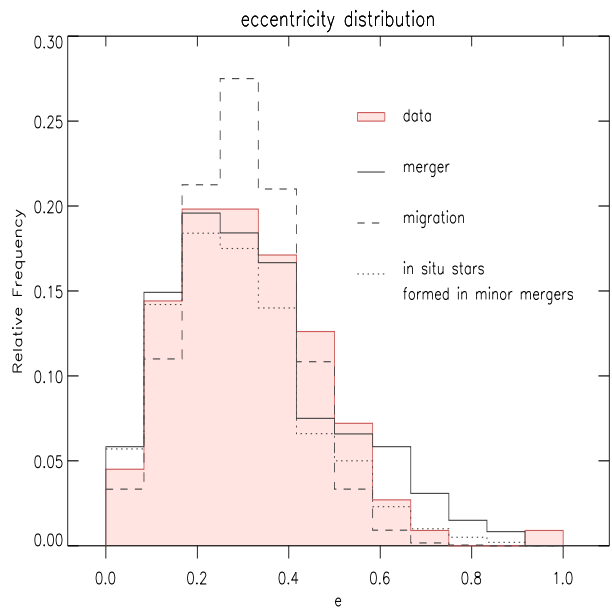}
\caption{The eccentricity distribution of our 110 thick disk stars (red filled histogram). Overplotted are the predictions from two thick disk formation scenarios:
radial migration (dashed line) and minor gas-rich mergers (solid line). The subset of stars that formed during minor mergers \textit{in situ} is shown as the dotted
line.}\label{fig:f5}
\end{figure}

\subsubsection{[$\alpha$/Fe] abundances of the thick disk stars}
Further clues about the formation of the thick disk could possibly be gained 
by investigating $\alpha$ element abundances; the $\alpha$ elements provide important clues about the history (and in particular the duration) of star formation based on the different explosion timescales and enrichment products of supernovae II and Ia. Furthermore, the scatter in [$\alpha$/Fe] among thick disk stars gives clues about the mixing state of the inter-stellar medium (ISM) from which these stars formed. However, as already stated above,
RAVE's measurements of [$\alpha$/Fe] are not reliable for individual stars.
We nevertheless have estimated mean values for both disk components, because star-to-star
variations in [$\alpha$/Fe] are not random, so that we at least expect to observe a correct trend.

As can be seen from Table~\ref{tab:t4}, the thick disk stars have $\langle[\alpha/\text{Fe}]\rangle = 0.19\pm0.11$, 
slightly higher than thin disk stars with $\langle[\alpha/\text{Fe}]\rangle = 0.12\pm0.07$. 
Despite the large uncertainties, the mean [$\alpha$/Fe] values of thick and thin disks 
are quite similar to the values given in Fig.~1.4 of \citet{Nissen04}. Recently, \citet{ruc10} derived separate abundances of four $\alpha$ elements (Mg, Si, Ca, Ti) and iron for 243 metal-poor ([Fe/H]$<-$0.5) stars from RAVE of which $\sim40\%$ have been classified as thin or thick disk stars based on their positions and velocities. These authors found $[\alpha/\text{Fe}]$ to be enhanced in thick disk stars with low scatter over a wide range of [Fe/H], although no mean values are given. This would imply rapid star formation on a timescale $\lesssim1$ Gyr and a high degree of mixing of the gas from which these stars formed; this in principal is consistent with the \texttt{in situ} formation of the thick disk from accreted gas, although such a mechanism would predict further stars that have been directly accreted at later times and hence posess lower [$\alpha$/Fe] values, but have not been observed by \citet{ruc10}. We note already that our sample contains such ``missing'' stars at low metallicities, but due to their small number and the great uncertainties on our $[\alpha/\text{Fe}]$ values, we make no further comparison to the sample of \citet{ruc10}.

\subsection{Derivation of the scale length of thick disk\label{sec:s43}}
From our estimates for the asymmetric drift, $V_\text{lag}$, and the components of the velocity ellipsoid, we are able to
estimate the scale length of the thick disk from the radial component of the steady-state Jeans equation:
\begin{equation}\label{eq:Jeans1}
\frac{R}{\rho}\frac{\partial(\rho\langle v_R^2\rangle)}{\partial R}+\frac{R}{\rho}\frac{\partial(\rho\langle v_Rv_z\rangle)}{\partial z}+\langle v_R^2\rangle-\langle v_\phi^2\rangle+R\frac{\partial\Phi_\text{tot}}{\partial R}=0
\end{equation}
Here, $v_R$, $v_\phi$ and $v_z$ denote Galactocentric cylindrical velocity components, $\rho(R,z)$ is the volume density of the thick disk stars, i.e. the tracer population, and $\Phi_\text{tot}(R,z)$ the total gravitational potential of the Galaxy.
We assume that our thick disk stars represent a relaxed population of stars in equilibrium with the total gravitational potential $\Phi_\text{tot}(R,z)$, which is given through equation~\eqref{eq:potential}. Close to the midplane at the sun's position, this potential is dominated by the thin disk\footnote{The density of the Miyamoto-Nagai disk at $R_\odot=8.5$ kpc is 0.16 $M_\odot$ pc$^-{3}$, which is 284 times larger than the contribution from the bulge and halo.}, which justifies our choice of a single disk component \citep[see also Sect.~4.1 in][]{gir06}. On top of that, we are free to choose an independent description for the thick disk density, for which we assume an exponential profile, because it allows an easy solution of the Jeans equation.

We evaluate the radial Jeans equation for a steady state disk at $R=R_\odot$ and $z=0$. This is justified by the small distances that our stars probe ($z<700$ pc, $\vert R-R_\odot\vert<950$ pc). Thus we can replace the term $R\frac{\partial\Phi_\text{tot}}{\partial R}$ in equation~\eqref{eq:Jeans1} with $v_c^2$, the squared circular velocity at $R_\odot$. 

Usually, one assumes that there is no net radial and vertical stellar motion in the solar neighborhood. This assumption has recently been challenged by \citet{sie11}, who found a radial velocity gradient of at least 3 km s$^{-1}$ kpc$^{-1}$ towards the Galactic center from a sample of 213,713 RAVE stars with $\vert z\vert<1$ kpc. However, before this result is confirmed by more studies, we will here keep the conservative approach and use $\langle v_R\rangle=0$ and $\langle v_z\rangle=0$, so that the term $\langle v_R^2\rangle$ can be replaced by $\sigma_R^2$ and $\langle v_rv_z\rangle$ by $\sigma_{Rz}$. We further have $\frac{\partial\rho}{\partial z}=0$ at $z=0$ by symmetry. With $\langle v_\phi^2\rangle=\sigma_\phi^2+\langle v_\phi\rangle^2$, the radial Jeans equation then reads
\begin{equation}\label{eq:Jeans2}\begin{split}
v_c^2-\langle v_\phi\rangle^2=&\sigma_\phi^2-\sigma_R^2-\frac{R}{\rho}\frac{\partial(\rho\sigma_R^2)}{\partial R}-R\frac{\partial\sigma_{Rz}}{\partial z}\\
=&\sigma_R^2\Bigl(\frac{\sigma_\phi^2}{\sigma_R^2}-1-\frac{R}{\rho\sigma_R^2}\frac{\partial(\rho\sigma_R^2)}{\partial R}-\frac{R}{\sigma_R^2}\frac{\partial\sigma_{Rz}}{\partial z}\Bigr).
\end{split}
\end{equation}
The cross term $\frac{\partial\,\sigma_{Rz}}{\partial\,z}$ vanishes if the principal axes of the velocity ellipsoid are aligned with the cylindrical coordinate system; our data indicate that this is not strictly the case. We measure a tilt angle between the long axis of the velocity ellipsoid and the sun-center line of $\alpha_{Rz}=5.6^\circ\pm0.4^\circ$, where $\alpha_{Rz}$ is given by
\begin{equation}
\tan(2\alpha_{Rz}) = \frac{\sigma_{Rz}^2}{\sigma_R^2-\sigma_z^2},
\end{equation}
and the uncertainty has been evaluated by computing 1000 Monte Carlo values using the velocity uncertainties for individual stars.
Our measurement means that the principal long axis of the velocity ellipsoid points slightly above the Galactic center.
This angle is small enough, however, to neglect it's influence on our evaluation close to the midplane. We therefore neglect the cross term in equation~\eqref{eq:Jeans2}.

The term $-\frac{\partial(\rho\sigma_R^2)}{\partial R}$ can be simplified by assuming that the shape of the thick disk's velocity ellipsoid is not varying with respect to $R$, an assumption that is reasonable given the small spatial extend of our data. This implies $\sigma_R^2\propto\sigma_z^2$. If the thick disk would be self-gravitating, the squared vertical velocity dispersion would be proportional to the thick disk density. This would imply \citep[e.g.][]{gir06}
\begin{equation}\label{eq:Jeans3}
-\frac{R}{\rho\sigma_R^2}\frac{\partial (\rho\sigma_R^2)}{\partial R}=\frac{2R}{h_R}.
\end{equation}
Here, the density distribution of the thick disk is given by $\rho(R,0)=\rho_0\exp(-R/h_R)$.
However, above we have stated that our thick disk stars are presumably dominated by the thin disk's gravitational potential; therefore it seems more reasonable to assume that the squared vertical velocity dispersion of the \textit{thick disk} stars is proportional to the density of the \textit{thin disk}. We thus have $\sigma_R^2\propto\sigma_z^2\propto\rho_\text{thin}(R,0)$, where $\rho_\text{thin}$ is given analytically by the Miyamoto-Nagai disk density \citep{miy75}. It then follows that
\begin{equation}\label{eq:Jeans4}
-\frac{R}{\rho\sigma_R^2}\frac{\partial (\rho\sigma_R^2)}{\partial R}=\frac{R}{h_R}-\frac{R}{\rho_\text{thin}}\frac{\partial\rho_\text{thin}}{\partial R}.
\end{equation}
Equations~\eqref{eq:Jeans3} and \eqref{eq:Jeans4} represent two extreme cases of a totally non-self-gravitating and a totally self-gravitating thick disk, respectively. The truth presumably lies somewhere between these two cases \citep{gir06}, but we will first use equation~\eqref{eq:Jeans4} for self-consistency. We will also investigate, however, how our estimate of the thick disk scale length changes if we apply equation~\eqref{eq:Jeans3} instead (i.e. if we assume a self-gravitating thick disk).

We now evaluate equation~\eqref{eq:Jeans2} with our measurements and assumptions to obtain an estimate for $h_R$, the radial scale length of the thick disk. Setting $\sigma_R=\sigma_U$, $\sigma_\phi=\sigma_V$ and $v_c-\langle v_\phi\rangle=V_\text{lag}$, equation~\eqref{eq:Jeans2} yields
\begin{equation}\label{eq:Jeans5}
2v_cV_\text{lag} - V^2_\text{lag} = \sigma^2_{U}\Bigl(\frac{\sigma^2_{V}}{\sigma^2_{U}}-1-\frac{R}{\rho_\text{thin}}\frac{\partial\rho_\text{thin}}{\partial R}+\frac{R}{h_R} \Bigr).
\end{equation}
$v_c$ is related to the Galactic potential through $v_c^2=R\frac{\partial\,\Phi_\text{tot}}{\partial\,R}$. The potential \eqref{eq:potential} that we used to integrate the orbits yields a circular velocity at the solar radius $R_\odot$ = 8.5 kpc of 223.1 km s$^{-1}$, not too different from our adopted LSR velocity of 220 km s$^{-1}$. 

With the values for our thick disk velocity ellipsoid, $(\sigma_{U}, \sigma_{V}, \langle v_\phi\rangle)$ = (57, 36, 163)
km\,s$^{-1}$, and our adopted
value for the the solar radius, $R_\odot$ = 8.5 kpc, 
we estimate the scale length of the thick disk as 
\begin{equation}
h_{R}= 1.5\pm0.3\,\text{kpc},
\end{equation}
where the uncertainty is determined by 1000 Monte Carlo computations of $h_R$, each time adding a normally distributed random error to the observables with standard deviation given by their uncertainties in Table~\ref{tab:t4}.
Using our adopted LSR velocity, $V_\text{LSR}=220$ km s$^{-1}$, instead of $v_c$ changes this result slightly to 1.6$\pm$0.3 kpc.

Our estimated value seems too small compared to scale length estimates from the literature that typically range from 2.2--3.6 kpc \citep{Morrisonetal90,Soubiran93,Robinetal,Ojhaetal,val06,jur08,car10}. Our estimate could be biased by some of the assumptions we made. One drawback is that the gravitational potential that dominates the motion of the tracer stars has to be globally well-tuned. This follows directly from equation~\eqref{eq:Jeans5}, which depends mainly on the circular velocity and the measured velocity dispersions. For example, by doubling both the scale parameter $a$ and the bulge mass $M_\text{bulge}$ in our total potential \eqref{eq:potential}, and -- not unrealistically -- assuming the same observations, we would obtain similar values for the radial velocity and scale length (219 km s$^{-1}$ and 1.3$\pm$0.2 kpc, respectively), despite having created an overall different galaxy with the local stellar density reduced by 34\%. As another example, if we assume the thick disk to be self-gravitating, i.e. using equation~\eqref{eq:Jeans3}, our result would change to $h_R=(2.2\pm0.3)$ kpc. Although this seems a less likely assumption, \citet{gir06} also found that their best thick disk models where those that assumed the ``self-gravitating form of the pressure term for the thick disk''.

We therefore conclude that our data together with the adopted Galaxy model and assumptions outlined above allow us to derive a rough estimate for the thick disk scale length in the range $1.5-2.2$ kpc, which is at the lower end of other values estimated so far.

\subsection{The metal abundance gradient\label{sec:s44}}
The relation between the maximal height above or below the galactic plane that a star reaches, $z_\text{max}$,
and the metal abundance for our thick disk stars
is shown in Fig.~6. No vertical abundance gradient is evident based on
the mean values of [M/H] and $z_\text{max}$ (filled dots) in Fig.~6.

\begin{figure}
\centering
\includegraphics*[width = 10cm, height = 6cm]{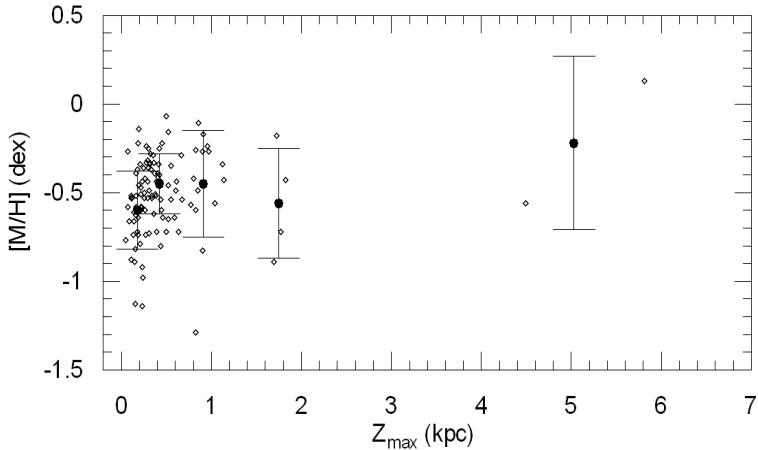}
\caption{[M/H] versus $z_\text{max}$  
for thick disk stars with $-22 < X < -5$.}\label{fig:f6}
\end{figure}

We also investigate the evidence for a radial metallicity gradient in the thin disk by using
the mean orbital radii, $R_m$, that are fairly robust against stellar migration \citep{Grenon87}.
In Fig.~7, we show the trend of [M/H] versus $R_{m}$ for our thin disk stars; the fitted radial gradient is 
$-0.07\pm$0.01 dex kpc$^{-1}$. \citet{Nordstrometal} found
a radial iron abundance gradient of $-$0.09 dex kpc$^{-1}$ for stars younger than 10 Gyr.
Our estimate is also consistent with a gradient
of $-$0.10 dex kpc$^{-1}$ predicted
by the chemical evolution model of \citet{sch09} and – qualitatively – with the gradient expected for stars close to the plane (and hence young) from simulations carried out by \citet[][Fig.2]{ros08}, although the latter authors stated no values. In both of these simulations, the thick disk of the Galaxy is explained naturally through stellar migration, leading to a net outward and inward motion of stars with time \citep[see also][]{loe10}. In contrast, the chemical evolution models of \citet{chi97} and \citet{chi01}, where the disk is decomposed into radial annuli that neither exchange gas nor stars, predict a much lower radial gradient of the order $\sim0.01-0.03$ dex kpc$^{-1}$ \citep[][Table~4]{chi97}. In these ``Two Infall'' models, the thin disk slowly forms inside-out after the thick disk and halo from a separate gas infall epoch. A gas density threshold for star formation is applied, preventing the growth of abundances and the build-up of a steeper metallicity gradient in the outer regions of the disk (including the solar neighborhood), where the gas density is low. Thus it seems that radial migration models are better suited to account for the observed radial metallicity gradient than disk formation scenarios where stars are basically confined to their radial birth places.

\begin{figure}
\centering
\includegraphics*[width = 12cm, height = 7cm]{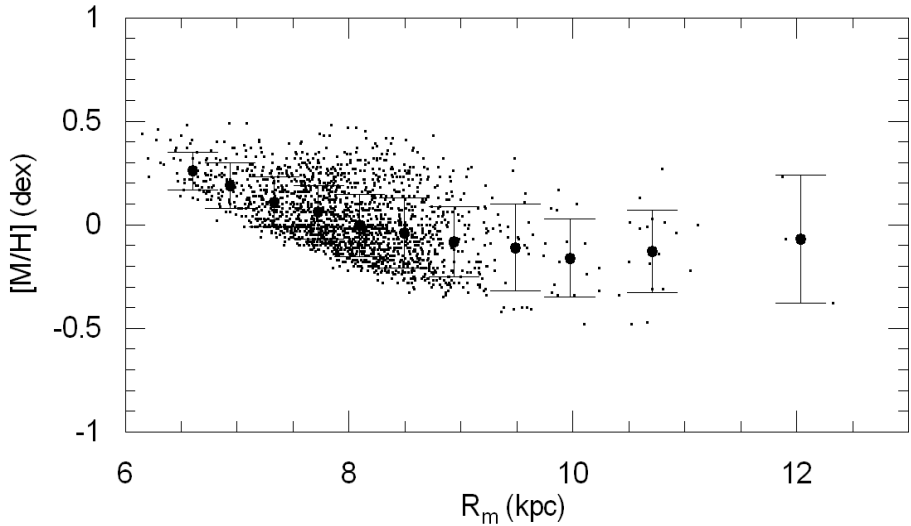}
\caption{[M/H] versus the mean Galactocentric distance $R_{m}$ for thin disk stars with $X\leq-33$.}\label{fig:f7}
\end{figure}

\section{Conclusions\label{sec:s5}}

Based on a sample of 4027 putative main sequence stars selected from the RAVE second data release
which have distance estimates taken from \citet{bre10}, 
we successfully reproduce the mean kinematic properties of the thick and 
thin disk populations compared to previous
results from the literature. The thick disk dominates our sample for metallicities [M/H]$\lesssim-0.8$ (Fig.~3), which explains the rapid rise in the vertical velocity dispersion occurring at this metallicity (Fig.~2a). On the other hand, we observe only a weak rise of $\sigma_W$ in the range $-0.50 < \text{[M/H]} < +0.50$; this rise of $\sigma_W$ could be explained either by the growing fraction of thick disk stars or heating of old thin disk stars or both. Contribution of the heating effect would imply that there is a trend in the thin disk's age-metallicity relation, namely that older stars are being on average more metal-poor. The literature on the age-metallicity relation of the thin disk has yielded mixed results, with some data showing a clear trend \citep[e.g.][]{Edvardssonetal93}, while others not \citep{hol07}, and it seems that selection effects in the different surveys are responsible for these differences \citep{hol07}. A study of the age-metallicity relation with RAVE stars is currently undertaken by \citet{ang09}, and the reader is referred to their poster for a short review and outlook on this subject.

A separation between thin and thick disk is 
achieved by combining kinematics and metal abundances in the so-called 
$X$ stellar population parameter defined by \citet{sch93}.  
We have estimated the velocity ellipsoids of the thin and thick disk to be 
$(\sigma_{U}, \sigma_{V}, \sigma_{W})$ = (34, 19, 18) km s$^{-1}$ and (57, 36, 35)
km s$^{-1}$, respectively. The vertical velocity dispersion $\sigma_{W}= 35$\,km s$^{-1}$ 
for the thick disk is in good agreement with recent literature values in the range 
32--39 km s$^{-1}$ (Table~\ref{tab:t5}). It allows us to derive a thick disk scale length
in the range of $h_{R}= 1.5-2.2$ kpc, although this has to be considered a rough estimate, because the global form of the gravitational potential that dominates the motion of the thick disk stars is not constrained well.

There are four principal formation mechanisms for the thick disk: 
(i) a puffed-up (``heated'') thick disk resulting from the collision of 
a satellite with a pre-existing thin disk \citep{Quinnetal,Velazquezetal}; 
(ii) radial migration \citep{sch09};
(iii) formation of thick disk stars out of accreted gas during a gas-rich merger \citep{bro04}; 
(iv) direct accretion of stars into the disk \citep{Abadietal}.
\citet{Salesetal} have compared the eccentricity distribution of thick 
disk stars resulting from each of these four processes and found that
a prominent peak at low eccentricity is expected for the heating, migration 
and gas-rich merging scenarios, while the accretion scenario predicts
a large number of stars on highly eccentric orbits. As can be seen from 
a comparison between our Fig.~5 and Fig.~3 of \citet{Salesetal}, 
our eccentricity distribution centered at
$\langle e \rangle = 0.31\pm0.16$ matches a scenario in which gas-rich minor mergers and/or radial 
migration are the key drivers of thick disk formation. We point out, however, that our stars and
the stars displayed in \citet{Salesetal} occupy different heights above and below the plane.
It is expected that the number of stars on highly eccentric orbits increases with $\vert z\vert$, broadening 
the high-eccentricity tail of the $e$-distribution. This would make
radial migration more unlikely than minor mergers because the former predicts a very narrow peak at low $e$ and
lacks a high-eccentricity tail. However, our radial metal abundance gradient of $-0.07$ dex kpc$^{-1}$ is consistent 
with the chemical evolution model of \citet{sch09},
in which radial migration alone is sufficient to explain 
the chemo-kinematic separation between thin and thick disks. We caution again that the
stars in our sample do not cover a sufficiently large part of the thick disk (in particular in $z_\text{max}$)
in order to draw definite conclusions on the formation of the thick disk as a whole. In addition, it would be interesting
to compare other diagnostics of the different models, e.g. radial or vertical metallicity gradients.

We have not detected any vertical metallicity gradient in the thick disk, 
while radial mixing would predict [M/H] to fall with increasing $z$ due to a net upward motion of stars over time \citep{loe10}; 
however, to what extend strongly depends on the enrichment history of the Galaxy 
and on the undetermined inside-out formation (Ralph Sch{\"o}nrich, 2010, private communication).
In addition, our data are sparse at
large $z$ and exhibit a strong dispersion. This is a general problem 
in the literature and the main reason that
no consensus on a vertical metallicity gradient exists yet. 
In fact, the large dispersion observed in all data sets so far
may reflect the intrinsic complexity of Galactic evolution including 
processes such as in-fall, out-flow, radial mixing 
and stellar interactions \citep{Cuietal}.

Our value for the radial metallicity gradient of the thin disk, on the other hand, is consistent with estimates from the Geneva-Copenhagen survey \citep{hol07} and predictions made by radial migration models \citep{sch09}. These gradients are steeper than expected for the chemical evolution models of \citet{chi97,chi01}, who assume a gas density threshold for star formation and a gradually declining inside-out growth of the Galactic thin disk. Thus it seems that radial migration to some extend is needed to account for the observed gradient.

In this study we have used the uncalibrated metallicities given in the RAVE DR2 catalog. When using the calibrated ones instead, but keeping the definition of the X parameter,
we only end up with 34 thick disk stars. This is expected, because the calibration of the metallicity parameters that are derived in the RAVE pipeline through a penalized $\chi^2$ comparison with a grid of synthetic spectra shifts these on average to higher values \citep[see Fig.~17 in][]{zwi08}. In our case, the mean [M/H] of the thick disk stars changes from $-0.51$ to $-0.28$ when using calibrated [M/H] estimates (see Table~\ref{tab:t6}). The calibration mostly affects metal-poor stars, so that our estimate for the radial metallicity gradient in the thin disk changes from $-0.07\pm0.01$ dex kpc$^{-1}$ to $-0.04\pm$0.01 dex kpc$^{-1}$, which in principal is consistent with the ``Two-Infall'' model of \citet{chi97,chi01}. We also note a rise in the velocity dispersions and a drop in the rotational velocity of the thick disk from 163$\pm2$ to 132$\pm$8 km s$^{-1}$, while the results for the thin disk remain stable within the uncertainties. $\sigma_U$ and $\sigma_V$ generally show much less agreement between different studies in the literature (see Table~\ref{tab:t5}), so that we can not infer whether the values obtained with the calibrated [M/H] estimates are more reliable than the ones obtained with the uncalibrated metallicities. Inserting the values of $(\sigma_U,\sigma_V,V_\text{lag})$ from Table~\ref{tab:t6} into equation~\eqref{eq:Jeans5}, we obtain a radial scale length of the thick disk of (2.3$\pm$0.9) kpc, which is larger than our result for the uncalibrated metallicities and somewhat more consistent with the literature values cited in Section~\ref{sec:s4}. $\sigma_W$, however, is more consistent with recent literature estimates (Table~\ref{tab:t5}), which are all $<40$ km s$^{-1}$, when we adopt the uncalibrated metallicities as done in this study. With the calibrated [M/H], the mean eccentricity of the thick disk stars rises from 0.31$\pm$0.16 to 0.45$\pm$0.21, which would speak even more strongly against a formation of the thick disk by radial migration alone, although the uncertainties are quite large due to the small number of stars. We conclude that by using the calibrated [M/H] values given in the RAVE catalog, we would end up with fewer thick disk stars, but our main results would not change significantly within the uncertainties, apart from a change of the thick disk velocity ellipsoid from $(\sigma_U,\sigma_V,\sigma_W)$=(57$\pm$4,36$\pm2$,35$\pm$2) to (72$\pm$9,47$\pm$6,49$\pm$6) and an expected decrease in the radial metallicity gradient of the thin disk.

\begin{table*}
\tiny
\caption{Same as Table~\ref{tab:t4}, but now the stars have been classified into thin and thick disk members by using the calibrated metallicities. Note that [M/H] now refers to the calibrated metallicity.}\label{tab:t6}
\begin{tabular}{cccccccccc}
\hline
 X & N & $\langle[M/H]\rangle$ & $\sigma_{U}$ & $\sigma_{V}$ & $\sigma_{W}$ & $\langle V_{rot}\rangle$ & $\langle V_{\Phi}\rangle$ & 
 $\langle e \rangle$ & $\langle[\alpha/Fe]\rangle$\\
  &  & (dex) &km s$^{-1}$&  km s$^{-1}$& km s$^{-1}$&  km s$^{-1}$& km s$^{-1}$& & (dex)  \\

\hline
 $-22<X<-5$      & 34 &$-$0.28$\pm$0.25 & 72$\pm$9 & 47$\pm$6   & 49$\pm$6   &66$\pm$15  & 132$\pm$8   &  0.45$\pm$0.21 &0.12$\pm$0.12 \\
 $X\leq-33$      &3216 &+0.12$\pm$0.16   & 35$\pm$0.44 & 20$\pm$0.25 & 18$\pm$0.22 &187$\pm$1   & 220$\pm$0.35 &  0.12$\pm$0.07 &0.10$\pm$0.10 \\
\hline
\end{tabular} 
\end{table*}

\section{Acknowledgments}
We thank the anonymous referee for her/his comments and suggestions that helped to improve an older version of this manuscript.
Funding for RAVE has been provided by the Anglo-Australian Observatory, 
by the Astrophysical Institute Potsdam, by the Australian Research Council, 
by the German Research foundation, by the National Institute for Astrophysics at Padova,
by The Johns Hopkins University, by the Netherlands Research School for Astronomy, 
by the Natural Sciences and Engineering Research Council of Canada, 
by the Slovenian Research Agency, by the Swiss National Science Foundation, 
by the National Science Foundation of the USA (AST-0508996), 
by the Netherlands Organisation for Scientific Research, 
by the Particle Physics and Astronomy Research Council of the UK, by Opticon, 
by Strasbourg Observatory, and by the Universities of Basel, 
Cambridge, and Groningen. The RAVE web site is at www.rave-survey.org.

\clearpage

\end{document}